\documentstyle[sprocl]{article}

\input{psfig}

\def\hkpc{\ h^{-1}{\rm kpc}} 
\def\hMsun{\ h^{-1}M_{\odot}} 
\def\hMpc{\ h^{-1}{\rm Mpc}}

\def\be{\begin{equation}}
\def\ee{\end{equation}}
\def\bea{\begin{eqnarray}}
\def\eea{\end{eqnarray}}

\begin{document}


\title{SHAPES OF DARK MATTER HALOS}

\author{J. S. BULLOCK}

\address{Department of Astronomy, The Ohio State University , \\
140 W. 18th Ave, Columbus, OH 43210 
\\E-maIl: james@astronomy.ohio-state.edu} 

%

%


\maketitle\abstracts{ 
I present an  analysis of the density  shapes of dark  matter halos in
$\Lambda$CDM and $\Lambda$WDM  cosmologies.   The main  results 
are derived
from a statistical  sample of  galaxy-mass halos   drawn from a   high
resolution  $\Lambda$CDM  N-body   simulation.    Halo   shapes   show
significant trends with  mass and redshift:  low-mass halos are 
rounder than high mass halos, and, for a fixed mass, halos are 
rounder at  low  $z$.  Contrary  to previous expectations,  
which were based on
cluster-mass  halos  and non-COBE normalized simulations, $\Lambda$CDM
galaxy-mass halos  at $z=0$ are not strongly flattened,  with short to
long axis ratios of $s = 0.70 \pm 0.17$.  I go on  to study how the
shapes of  individual   halos  change  when going   from   a $\Lambda$CDM
simulation  to a  simulation with  a  warm dark matter power  spectrum
($\Lambda$WDM).  
Four halos were compared, and,  
on average, the  WDM halos are more spherical
than their CDM counterparts ($s\simeq  0.77$ compared to $s   \simeq
0.71$).  A larger sample  of objects will be needed
to test whether the trend is significant.}

\section{Introduction}

A   variety of  observational indicators suggest that galaxies are
embedded within massive,  extended dark matter halos,  lending support
to the   idea that a    hierarchical,  cold dark matter   (CDM)  based
cosmology  may provide a real description  of the universe, especially
on large scales  (see, e.g., the review by Primack~\cite{p:01} and 
references therein).  A
useful small-scale test of  CDM and variant   theories may come  from
observations aimed at inferring the density structure  of  dark 
halos.  Specifically,  the
quest to  measure dark   halo shapes  has  developed  into a rich  and
complex subfield  of its own.~\cite{these:01,s:99}
  
Predictions for shapes of dark  matter halos formed by dissipationless
gravitational  collapse  are  most reliably   derived using  numerical
studies.~\cite{f:88,dc:91,w:92}
These  past   investigations  focused  on galaxy-sized  halos
formed  from power-law or pre-COBE CDM  power spectra,  and found that
halos were typically flattened triaxial structures, with short-to-long
axis  ratios  of  $s  \simeq 0.5  \pm  0.15$.   Examinations  based on
currently favored  cosmologies have been done, but they
studied only cluster mass halos,~\cite{m:95,t:97,m:01} and also found significantly
flattened objects, $s \sim 0.4-0.5$.
 Interestingly, Thomas and collaborators~\cite{t:97}
saw an indication that low  matter density  models (with early
structure formation) tended  to  produce more spherical clusters  than
high density models.   This result  was qualitatively consistent  with
indications from  Warren et al.~\cite{w:92}  that high-mass halos are more
flattened than (early-forming)  low-mass   halos.  In light  of  these
hints that formation  history plays a  role in setting halo shapes, it
is important   to reexamine  the  question  for  a  currently standard
cosmological model.  Specifically,  this  work aims at  characterizing
halo  flattening as  a function of  mass  and redshift for the popular
flat  $\Lambda$CDM cosmology.  I also explore  how shapes of halos are
affected  when  going  to a $\Lambda$WDM  model,  in  which  the power
spectrum is damped on small scales.

\section{Simulations and shape measurement}

The simulations were performed using the Adaptive Refinement Tree
(ART)  code.~\cite{kkk:97}  The  main results are derived  from
a $\Lambda$CDM simulation with  $\Omega_m = 0.3, \Omega_{\Lambda} = 0.7,
h = 0.7,$  and $\sigma_8 = 1.0$, which  followed $256^3$ particles  of
mass $1.1 \times 10^9 \hMsun$ within a periodic box of comoving length
$60  \hMpc$, obtaining a  formal force  resolution of  $1.8 \hkpc$.
A second simulation was run
until $z=1.8$ with the same particle number, half the box size, and
thus eight times the mass resolution; it was used to check resolution
issues.
 Dark  halos  were identified  using a (spherical) bound density
maxima method,~\cite{b:01a} and masses were determined
by counting
particles within the spherical virial   radius $R_{\rm v}$, inside which  the
mean overdensity  has dropped to  a  value 
$\Delta_v \simeq (18 \pi^2 +  82p - 39p^2)/(1+p)$, 
where   $p \equiv \Omega_m(z)  - 1$.   This sample contains 
$\sim 800$ halos that span  the mass range 
$3 \times 10^{11} -   5 \times  10^{14}$.

A second pair of simulations~\cite{ar:01} 
were used to explore how shapes of
halos are affected by damping the power spectrum.  We compare 
four halos, each simulated from the same initial conditions, for a
$\Lambda$CDM  and $\Lambda$WDM model with the multiple-mass
ART code and the same cosmological parameters and effective
mass per particle discussed above.  The filter 
mass scale for the $\Lambda$WDM
run was $1.7\times 10^{14} \hMsun$, and the four halos we compare
have masses $(2-8) \times 10^{13} \hMsun$.~\footnote{
Although this filtering mass is much too large to be
an consistent with  with Ly-$\alpha$ forest measurements,~\cite{n:01} 
the simulation provides a useful comparison
to test the effect of an imposed filtering scale, since the
halo masses considered are well within the affected regime.  
Since the effect should simply scale with the filtering mass, 
similar results would be obtained for $\sim 5\times10^{10}\hMsun$ halos
formed from  $\sim 1$keV WDM particles.}  Simulation results were
kindly supplied by P. Col\'{\i}n.    

Halo axis ratios are determined
using the moments of the particle distributions within the virial
radius $R_{\rm v}$.   The
short-to-long axis ratio, $s$, and intermediate-to-long axis ratio $q$
are calculated by iteratively diagonalizing the tensor~\cite{dc:91}
\begin{eqnarray}
M_{ij} & = & \Sigma \frac{x_i x_j}{a^2}, \quad a \equiv \sqrt{x^2 + 
\frac{y^2}{q^2} + \frac{z^2}{s^2}},
\end{eqnarray}
where $q^2 \equiv M_{yy}/M_{xx}$ and $s^2 \equiv M_{zz}/M_{xx}$.

\begin{figure}
{\noindent \begin{minipage}[t]{2.3in} \centering 
 {\psfig{file=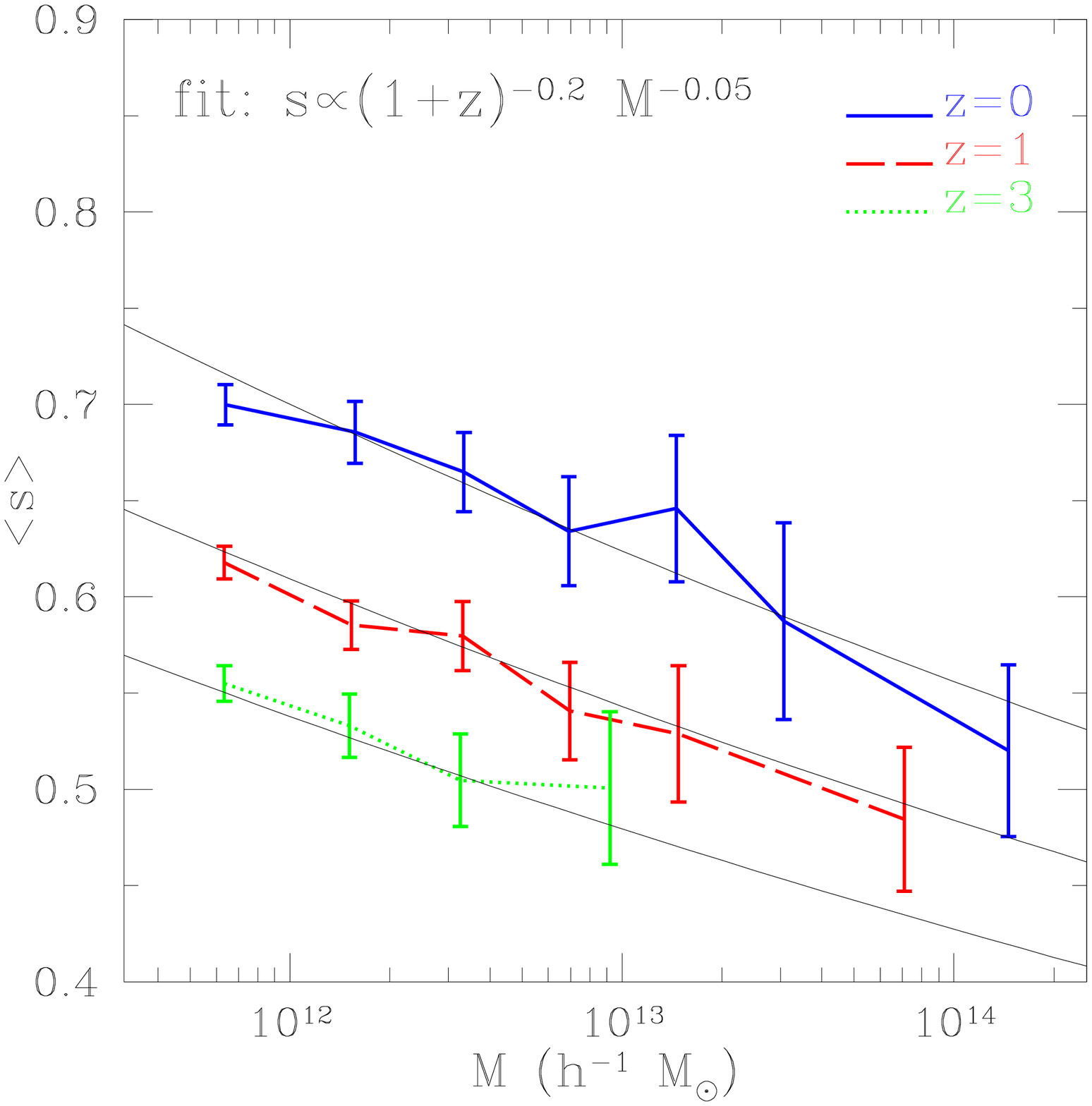,height=2.3in}} \end{minipage} \hfill
 \begin{minipage}[t]{2.3in} \centering 
 {\psfig{file=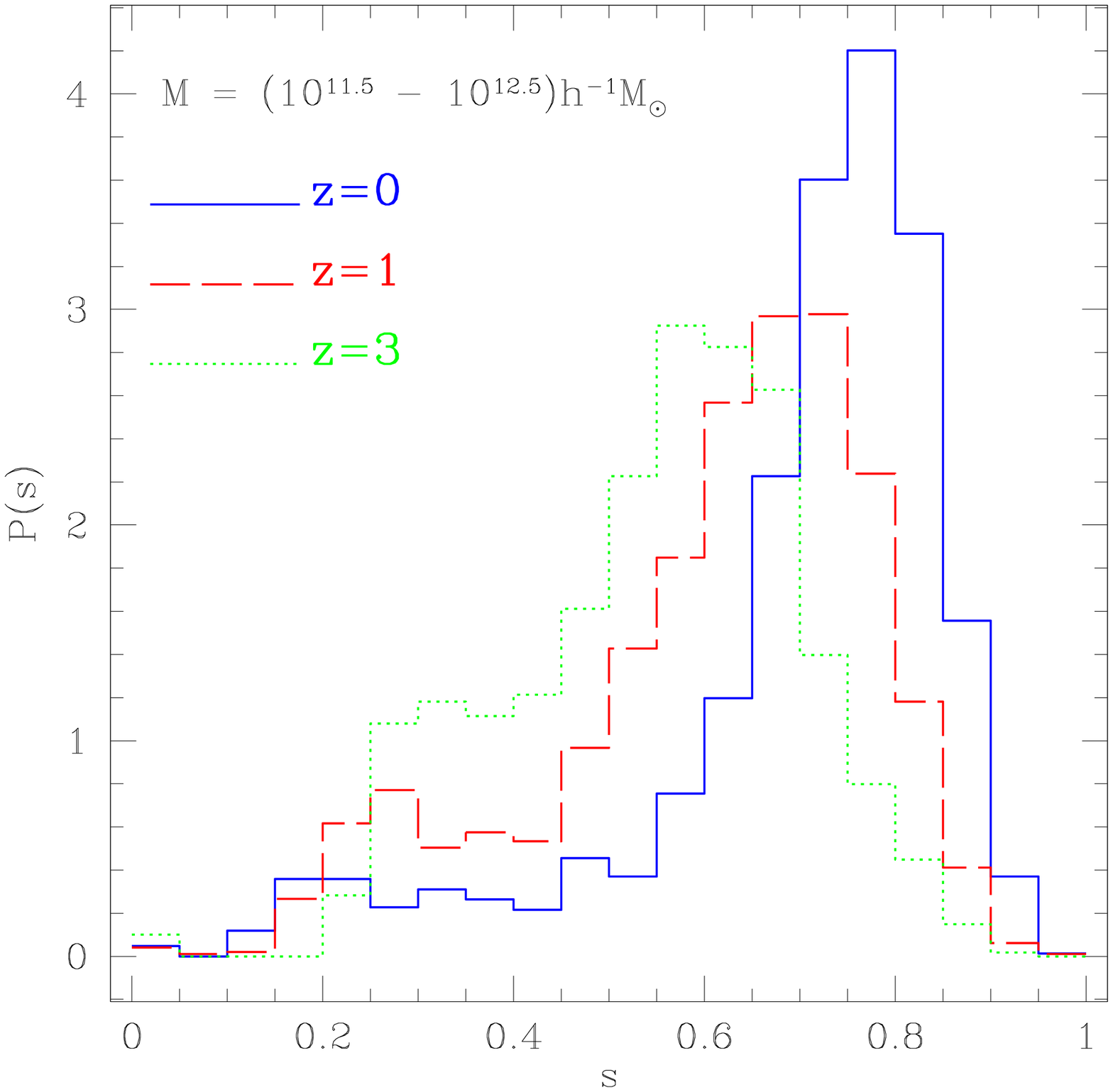,height=2.3in}} \end{minipage} }
\caption{ (Left) The average short-to-long axis ratio, $s$, as a
function of halo mass at $z=0$, $1$, and $3$. Error bars 
reflect the Poisson uncertainty associated with the number of halos in
each mass bin and \textit{not}
the scatter about the relation. (Right) Distribution
of measured $s$ values for $\sim 10^{12} \hMsun$ halos as a function
of $z$ }
\label{fig:galspectra}
\end{figure}

\section{Results}


The left  panel in  Figure 1  shows  the  average  value of  $s$  as a
function of halo  mass   for three redshifts, $z=0$, 1, and 3.   
 Low mass halos, on average, are rounder than high mass
halos, as are halos   of  fixed mass at   low  $z$.  The relation   is
well-approximated    by      $s     \simeq        0.7       (M/10^{12}
\hMsun)^{-0.05}(1+z)^{-0.2}$  over  the   mass   and redshift   ranges
explored.  The right  panel of Figure  1 shows the distribution of $s$
parameters for galaxy-mass halos as a function of z.  The  average
and  rms dispersion in these distributions   are 
$s =  0.70 \pm 0.17$, $0.61  \pm  0.17$, and $0.55 \pm   0.15$,
for $z=0$, 1, and 3, respectively.
 Note  that the distributions are quite    non-Gaussian, 
with a significant  tail   of highly flattened halos.~\footnote{
In order to check resolution effects, the
$z=3$ distribution was compared to the corresponding one obtained
in the high resolution simulation and found to be statistically
equivalent.}

How  do the shapes  of halos  change with  radius?   The axial  ratios
presented in Figure  1 were obtained using particles within $R_{\rm v}$. 
The left  panel of Figure  2  shows this average  ``virial'' flattening
measurement  as a function  of halo  mass  (at $z=0$)  compared with  the
average  $s$ measured within  a sphere of   radius $30 \hkpc$ for each
halo.   Although the difference is quite  small for the low mass halos
(since $30 \hkpc$ contains much of the halo mass), generally halos are
rounder  at small radii.  Interestingly,  
within this fixed  central
radius, halos typically have the same  flattening, $s \sim 0.7$,
independent of the halo mass.  

Finally in the right panel  of Figure 2, we compare the
measured $s$ and $q$ values of four halos, simulated
in  $\Lambda$CDM and $\Lambda$WDM.
There is a tendency for the halos to be rounder (approaching the upper
right corner) in WDM, although one  halo (designated by the
triangles) does become slightly flatter in WDM.  The  average  
flattening shifts from  $s  \simeq 0.71$ for CDM  to $0.77$ for WDM,
but it is difficult to make strong conclusions based on four halos.
Indeed, Moore~\cite{m:01}
has simulated a single halo using CDM and WDM power spectra and
finds a very similar shape for each.  
A larger sample of objects will be needed
to test for systematic trends.

\begin{figure}
{\noindent \begin{minipage}[t]{2.3in} \centering 
 {\psfig{file=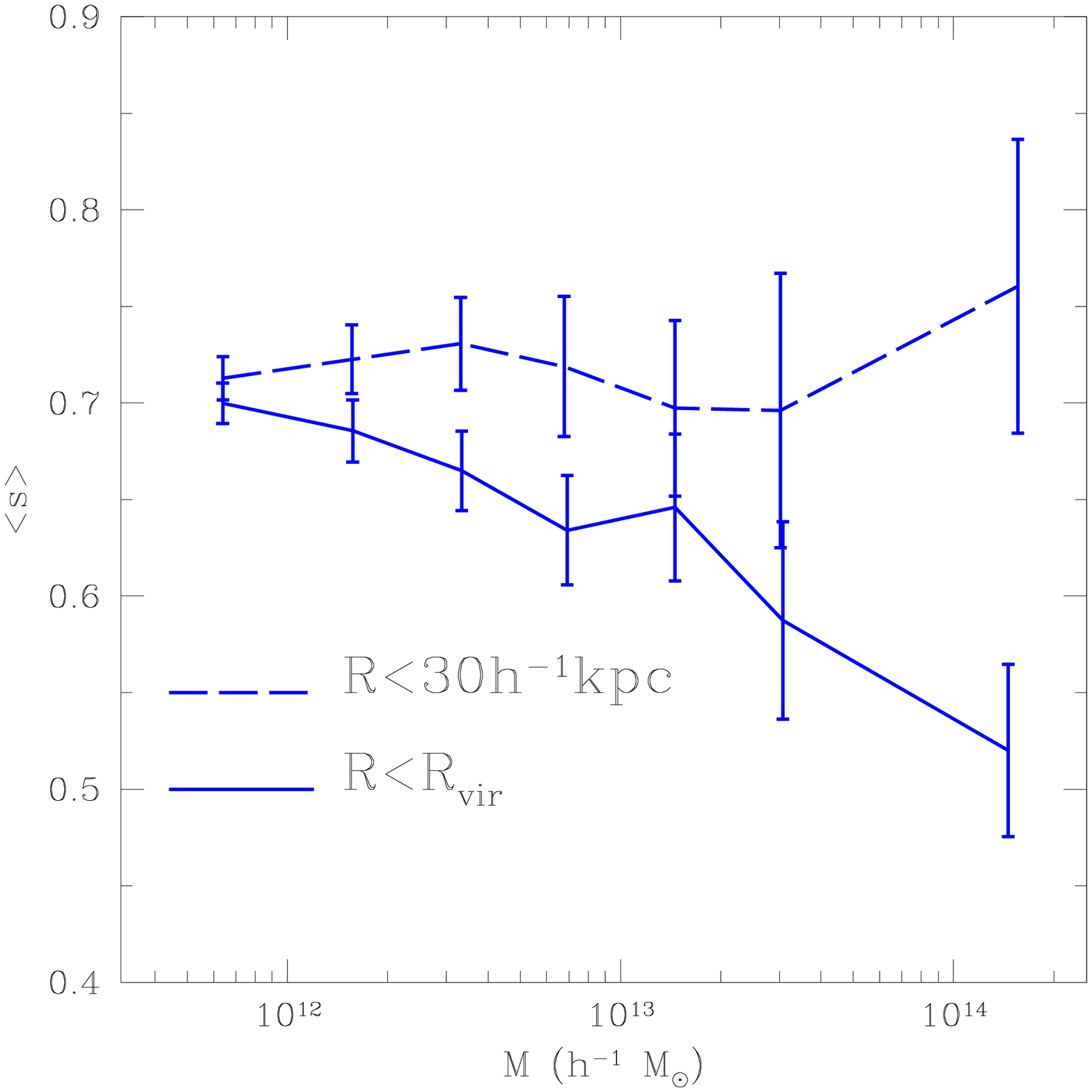,height=2.3in}} \end{minipage} \hfill
 \begin{minipage}[t]{2.3in} \centering 
 {\psfig{file=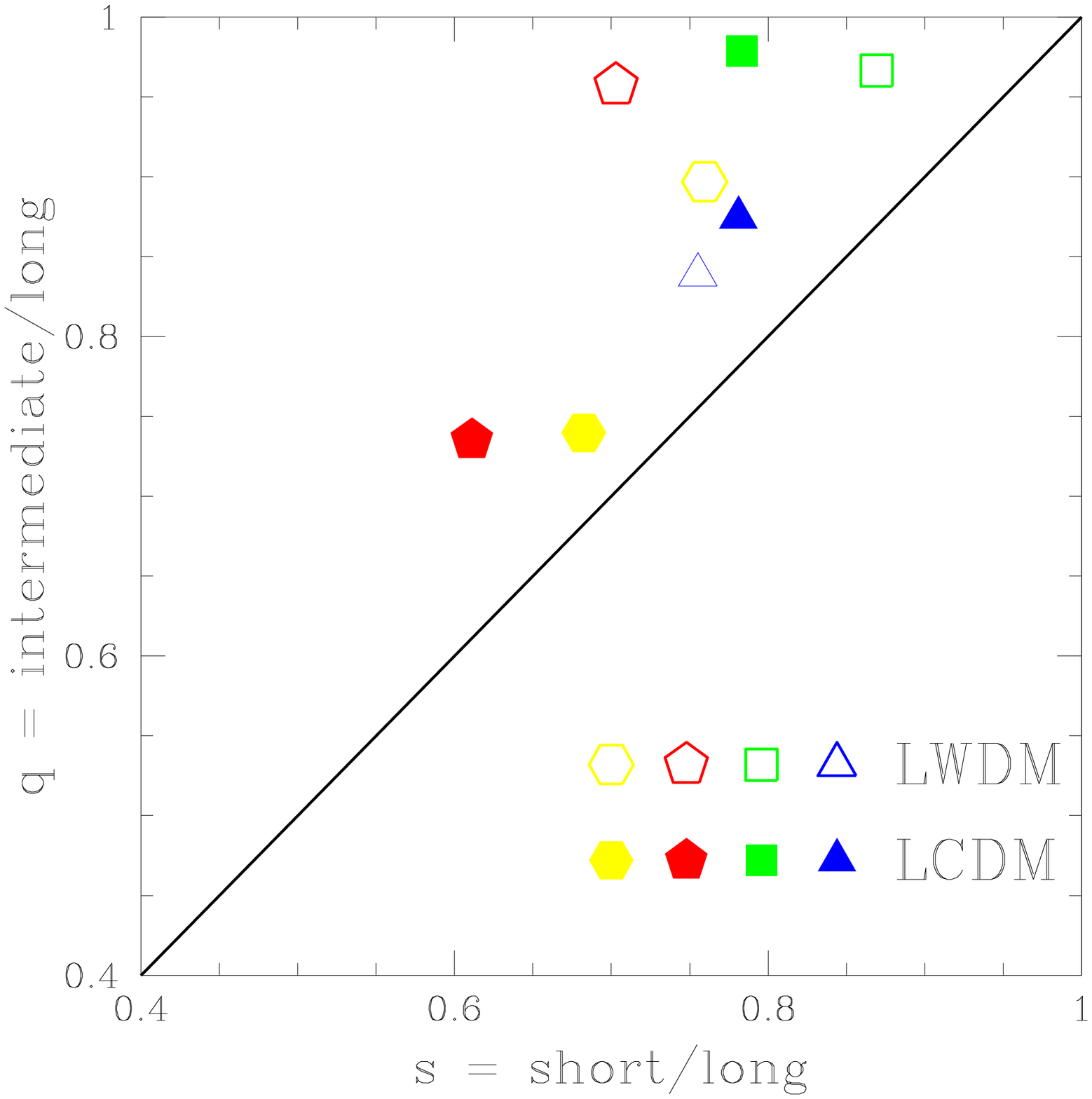,height=2.3in}} \end{minipage} }
\caption{(Left) $s$ as a function of mass measured within halo
virial radii $R_{\rm v}$ (solid) and within $30 \hkpc$ spheres from
halo centers. (Right) $q$ and $s$ parameters for halos simulated
using $\Lambda$CDM (solid symbols) and $\Lambda$WDM (open symbols)
cosmologies.  Individual halos, identified by mass and location
between the two runs maintain the same symbol shape.   
}
\label{fig:galspectra}
\end{figure}

\section{Conclusions}
We find that $\Lambda$CDM galaxy-mass halos at $z=0$ are
more spherical than previously believed, with $s = 0.70 \pm 0.17$.
High mass halos show more substantial flattening, as do halos
of fixed mass at high redshift: $s \propto (1+z)^{-0.2} M^{-0.05}$.
Halos are also more spherical in their centers, and tend to become
more flattened near the virial radius.  These trends suggest
collapsed structures become more spherical with time,
perhaps because they have had more time to phase mix and to obtain
isotropic orbit distributions.  It is also possible that the
accretion history itself plays a role. 
Halos formed within a $\Lambda$WDM simulation show a slight
indication of being less flattened than their $\Lambda$CDM 
counterparts.  This may be a reflection of substructure differences
between the two models,  but a larger number of halos will be
needed to decisively test this conclusion.  A more complete
description of these results, and some discussion of
shape correlation with other halo parameters is
presented in a forthcoming paper.~\cite{b:01b}

\section*{Acknowledgments}
I  thank  my  collaborators Pedro Col\'{\i}n,
Ricardo  Flores,   Andrey Kravtsov, Anatoly
Klypin, Ariyeh Maller, Joel Primack  and Risa Wechsler for allowing me
to  present  our results here.  Thanks to  Tsafrir Kolatt, Ben Moore,
and David Weinberg for insightful discussions, and
to Priya Natarajan for organizing this stimulating  meeting.  
This work was supported  by NASA  LTSA grant NAG  5-3525
and NSF grant AST-9802568.

\end{document}